\input harvmac
\input epsf
\def\journal#1&#2(#3){\unskip, \sl #1\ \bf #2 \rm(19#3) }
\def\andjournal#1&#2(#3){\sl #1~\bf #2 \rm (19#3) }

\def\frac#1#2{{#1\over#2}}

\def\inbar{\,\vrule height1.5ex width.4pt depth0pt}
\def\IC{\relax\hbox{$\inbar\kern-.3em{\rm C}$}}
\def\IR{\relax{\rm I\kern-.18em R}}
\def\IP{\relax{\rm I\kern-.18em P}}
\def\IZ{\relax{\rm I\kern-.18em Z}}

%
%

%
\catcode`\@=11
\def\slash#1{\mathord{\mathpalette\c@ncel{#1}}}
\overfullrule=0pt
\def\AA{{\cal A}}

\def\CC{{\cal C}}

\def\II{{\cal I}}

\def\LL{{\cal L}}

\def\SS{{\cal S}}

\def\VV{{\cal V}}

\def\underrel#1\over#2{\mathrel{\mathop{\kern\z@#1}\limits_{#2}}}

\catcode`\@=12


%

\def\det{{\rm det}}

\def \sinh{{\rm sinh}}
\def \cosh{{\rm cosh}}

\def\det{{\rm det}}


\def\p{{\partial}}


\def\unlockat{\catcode`\@=11}
\def\lockat{\catcode`\@=12}

\unlockat


\def\newsec#1{\global\advance\secno by1\message{(\the\secno. #1)}
\global\subsecno=0\global\subsubsecno=0\eqnres@t\noindent
{\bf\the\secno. #1}
\writetoca{{\secsym} {#1}}\par\nobreak\medskip\nobreak}
\global\newcount\subsecno \global\subsecno=0
\def\subsec#1{\global\advance\subsecno
by1\message{(\secsym\the\subsecno. #1)}
\ifnum\lastpenalty>9000\else\bigbreak\fi\global\subsubsecno=0
\noindent{\it\secsym\the\subsecno. #1}
\writetoca{\string\quad {\secsym\the\subsecno.} {#1}}
\par\nobreak\medskip\nobreak}
\global\newcount\subsubsecno \global\subsubsecno=0
\def\subsubsec#1{\global\advance\subsubsecno by1
\message{(\secsym\the\subsecno.\the\subsubsecno. #1)}
\ifnum\lastpenalty>9000\else\bigbreak\fi
\noindent\quad{\secsym\the\subsecno.\the\subsubsecno.}{#1}
\writetoca{\string\qquad{\secsym\the\subsecno.\the\subsubsecno.}{#1}}
\par\nobreak\medskip\nobreak}

\def\subsubseclab#1{\DefWarn#1\xdef
#1{\noexpand\hyperref{}{subsubsection}%
{\secsym\the\subsecno.\the\subsubsecno}%
{\secsym\the\subsecno.\the\subsubsecno}}%
\writedef{#1\leftbracket#1}\wrlabeL{#1=#1}}
\lockat


\def\TTT{{\cal{T}}}


\def\blobsix{{ \atop \epsfbox{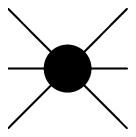}}}
\def\vertsix{{ \atop \epsfbox{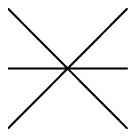}}}
\def\vertfour{{  \atop \epsfbox{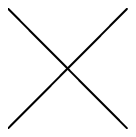}}}
\def\pmone{{ \atop \epsfbox{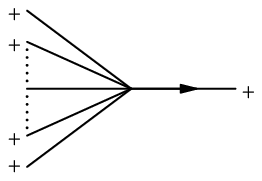}}}
\def\pmtwo{{ \atop \epsfbox{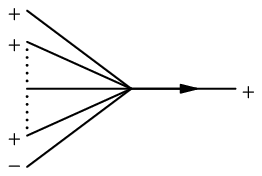}}}
\def\pmn{{  \atop \epsfbox{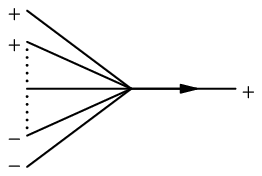}}}
\def\ppfone{{ \atop \epsfbox{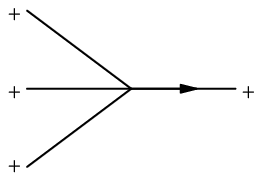}}}
\def\ppftwo{{  \atop \epsfbox{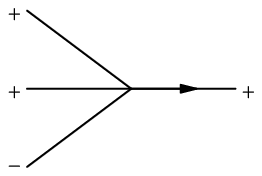}}}
\def\plainsix{{ \atop \epsfbox{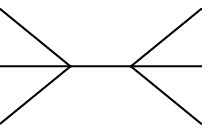}}}
\def\kkksix{{ \atop \epsfbox{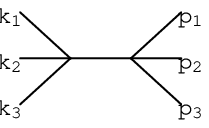}}}
\def\kkpsix{{ \atop \epsfbox{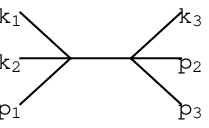}}}
\def\blobtwon{{ \atop \epsfbox{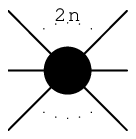}}}
\def\verttwon{{ \atop \epsfbox{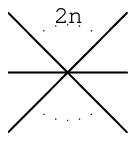}}}
\def\kkpkkp{{ \atop \epsfbox{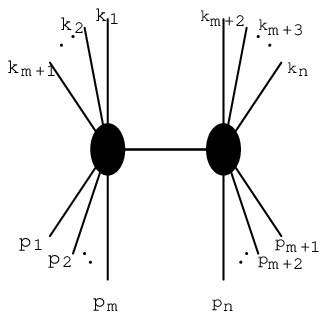}}}
\def\kkpcube{{ \atop \epsfbox{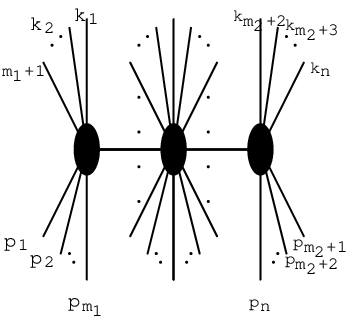}}}


\lref\SenTM{
A.~Sen,
``Dirac-Born-Infeld action on the tachyon kink and vortex,''
arXiv:hep-th/0303057.
}

\lref\SenQA{
A.~Sen,
``Time and tachyon,''
arXiv:hep-th/0209122.
}

\lref\SenAN{
A.~Sen,
``Field theory of tachyon matter,''
Mod.\ Phys.\ Lett.\ A {\bf 17}, 1797 (2002)
[arXiv:hep-th/0204143].
}

\lref\SenIN{
A.~Sen,
``Tachyon matter,''
JHEP {\bf 0207}, 065 (2002)
[arXiv:hep-th/0203265].
}

\lref\SenNU{
A.~Sen,
``Rolling tachyon,''
JHEP {\bf 0204}, 048 (2002)
[arXiv:hep-th/0203211].
}

\lref\LambertZR{
N.~Lambert, H.~Liu and J.~Maldacena,
``Closed strings from decaying D-branes,''
arXiv:hep-th/0303139.
}

\lref\LarsenWC{
F.~Larsen, A.~Naqvi and S.~Terashima,
``Rolling tachyons and decaying branes,''
JHEP {\bf 0302}, 039 (2003)
[arXiv:hep-th/0212248].
}

\lref\GarousiTR{
M.~R.~Garousi,
``Tachyon couplings on non-BPS D-branes and Dirac-Born-Infeld action,''
Nucl.\ Phys.\ B {\bf 584}, 284 (2000)
[arXiv:hep-th/0003122].
}

\lref\SenMD{
A.~Sen,
``Supersymmetric world-volume action for non-BPS D-branes,''
JHEP {\bf 9910}, 008 (1999)
[arXiv:hep-th/9909062].
}

\lref\LeblondDB{
F.~Leblond and A.~W.~Peet,
``SD-brane gravity fields and rolling tachyons,''
arXiv:hep-th/0303035.
}

\lref\BerkoozJE{
M.~Berkooz, B.~Craps, D.~Kutasov and G.~Rajesh,
``Comments on cosmological singularities in string theory,''
arXiv:hep-th/0212215.
}

\lref\KutasovAQ{
D.~Kutasov, M.~Marino and G.~W.~Moore,
``Remarks on tachyon condensation in superstring field theory,''
arXiv:hep-th/0010108.
}

\lref\KutasovQP{
D.~Kutasov, M.~Marino and G.~W.~Moore,
``Some exact results on tachyon condensation in string field theory,''
JHEP {\bf 0010}, 045 (2000)
[arXiv:hep-th/0009148].
}

\lref\HarveyNA{
J.~A.~Harvey, D.~Kutasov and E.~J.~Martinec,
``On the relevance of tachyons,''
arXiv:hep-th/0003101.
}

\lref\MarinoQC{
M.~Marino,
``On the BV formulation of boundary superstring field theory,''
JHEP {\bf 0106}, 059 (2001)
[arXiv:hep-th/0103089].
}

\lref\NiarchosSI{
V.~Niarchos and N.~Prezas,
``Boundary superstring field theory,''
Nucl.\ Phys.\ B {\bf 619}, 51 (2001)
[arXiv:hep-th/0103102].
}

\lref\WittenCR{
E.~Witten,
``Some computations in background independent off-shell string theory,''
Phys.\ Rev.\ D {\bf 47}, 3405 (1993)
[arXiv:hep-th/9210065].
}

\lref\WittenQY{
E.~Witten,
``On background independent open string field theory,''
Phys.\ Rev.\ D {\bf 46}, 5467 (1992)
[arXiv:hep-th/9208027].
}

\lref\GerasimovZP{
A.~A.~Gerasimov and S.~L.~Shatashvili,
``On exact tachyon potential in open string field theory,''
JHEP {\bf 0010}, 034 (2000)
[arXiv:hep-th/0009103].
}

\lref\SenMG{
A.~Sen,
``Non-BPS states and branes in string theory,''
arXiv:hep-th/9904207.
}

\lref\LambertHK{
N.~D.~Lambert and I.~Sachs,
``Tachyon dynamics and the effective action approximation,''
Phys.\ Rev.\ D {\bf 67}, 026005 (2003)
[arXiv:hep-th/0208217].
}

\lref\CallanMW{
C.~G.~Callan and I.~R.~Klebanov,
``Exact C = 1 boundary conformal field theories,''
Phys.\ Rev.\ Lett.\  {\bf 72}, 1968 (1994)
[arXiv:hep-th/9311092].
}

\lref\CallanUB{
C.~G.~Callan, I.~R.~Klebanov, A.~W.~Ludwig and J.~M.~Maldacena,
``Exact solution of a boundary conformal field theory,''
Nucl.\ Phys.\ B {\bf 422}, 417 (1994)
[arXiv:hep-th/9402113].
}

\lref\PolchinskiMY{
J.~Polchinski and L.~Thorlacius,
``Free Fermion Representation Of A Boundary Conformal Field Theory,''
Phys.\ Rev.\ D {\bf 50}, 622 (1994)
[arXiv:hep-th/9404008].
}

\lref\TseytlinMT{
A.~A.~Tseytlin,
``Sigma model approach to string theory effective actions with tachyons,''
J.\ Math.\ Phys.\  {\bf 42}, 2854 (2001)
[arXiv:hep-th/0011033].
}

\lref\TseytlinRR{
A.~A.~Tseytlin,
``Sigma Model Approach To String Theory,''
Int.\ J.\ Mod.\ Phys.\ A {\bf 4}, 1257 (1989).
}

\lref\FelderSV{
G.~N.~Felder, L.~Kofman and A.~Starobinsky,
``Caustics in tachyon matter and other Born-Infeld scalars,''
JHEP {\bf 0209}, 026 (2002)
[arXiv:hep-th/0208019].
}

\lref\DavidVM{
J.~R.~David, M.~Gutperle, M.~Headrick and S.~Minwalla,
``Closed string tachyon condensation on twisted circles,''
JHEP {\bf 0202}, 041 (2002)
[arXiv:hep-th/0111212].
}

\lref\BergshoeffDQ{
E.~A.~Bergshoeff, M.~de Roo, T.~C.~de Wit, E.~Eyras and S.~Panda,
``T-duality and actions for non-BPS D-branes,''
JHEP {\bf 0005}, 009 (2000)
[arXiv:hep-th/0003221].
}

\lref\KlusonIY{
J.~Kluson,
``Proposal for non-BPS D-brane action,''
Phys.\ Rev.\ D {\bf 62}, 126003 (2000)
[arXiv:hep-th/0004106].
}

\lref\GarousiPV{
M.~R.~Garousi,
``Off-shell extension of S-matrix elements and tachyonic effective actions,''
arXiv:hep-th/0303239.
}

\lref\MinahanTG{
J.~A.~Minahan and B.~Zwiebach,
``Gauge fields and fermions in tachyon effective field theories,''
JHEP {\bf 0102}, 034 (2001)
[arXiv:hep-th/0011226].
}

\lref\LambertFA{
N.~D.~Lambert and I.~Sachs,
``On higher derivative terms in tachyon effective actions,''
JHEP {\bf 0106}, 060 (2001)
[arXiv:hep-th/0104218].
}

\lref\TseytlinDJ{
A.~A.~Tseytlin,
``Born-Infeld action, supersymmetry and string theory,''
arXiv:hep-th/9908105.
}

\lref\GutperleXF{
M.~Gutperle and A.~Strominger,
``Timelike boundary Liouville theory,''
arXiv:hep-th/0301038.
}

\lref\GibbonsTV{
G.~Gibbons, K.~Hashimoto and P.~Yi,
``Tachyon condensates, Carrollian contraction of Lorentz group, and
fundamental strings,''
JHEP {\bf 0209}, 061 (2002)
[arXiv:hep-th/0209034].
}

\lref\SugimotoFP{
S.~Sugimoto and S.~Terashima,
``Tachyon matter in boundary string field theory,''
JHEP {\bf 0207}, 025 (2002)
[arXiv:hep-th/0205085].
}

\lref\MinahanIF{
J.~A.~Minahan,
``Rolling the tachyon in super BSFT,''
JHEP {\bf 0207}, 030 (2002)
[arXiv:hep-th/0205098].
}

\lref\IshidaFR{
A.~Ishida and S.~Uehara,
``Gauge fields on tachyon matter,''
Phys.\ Lett.\ B {\bf 544}, 353 (2002)
[arXiv:hep-th/0206102].
}

\lref\OhtaAC{
K.~Ohta and T.~Yokono,
``Gravitational approach to tachyon matter,''
Phys.\ Rev.\ D {\bf 66}, 125009 (2002)
[arXiv:hep-th/0207004].
}

\lref\GibbonsHF{
G.~W.~Gibbons, K.~Hori and P.~Yi,
``String fluid from unstable D-branes,''
Nucl.\ Phys.\ B {\bf 596}, 136 (2001)
[arXiv:hep-th/0009061].
}

\lref\KimHE{
C.~j.~Kim, H.~B.~Kim, Y.~b.~Kim and O.~K.~Kwon,
``Electromagnetic string fluid in rolling tachyon,''
JHEP {\bf 0303}, 008 (2003)
[arXiv:hep-th/0301076].
}

\lref\KimQZ{
C.~j.~Kim, H.~B.~Kim, Y.~b.~Kim and O.~K.~Kwon,
``Cosmology of rolling tachyon,''
arXiv:hep-th/0301142.
}

\lref\KutasovER{
D.~Kutasov and V.~Niarchos,
``Tachyon effective actions in open string theory,''
Nucl.\ Phys.\ B {\bf 666}, 56 (2003)
[arXiv:hep-th/0304045].
}

\lref\PolchinskiRR{
J.~Polchinski,
``String Theory. Vol. 2: Superstring Theory And Beyond,''
Cambridge, UK: Univ. Pr. (1998) 531 p.
}

\lref\SmedbackUR{
M.~Smedback,
``On effective actions for the bosonic tachyon,''
JHEP {\bf 0311}, 067 (2003)
[arXiv:hep-th/0310138].
}

\lref\DiFrancescoUD{
P.~Di Francesco and D.~Kutasov,
``World sheet and space-time physics in two-dimensional (Super)string
theory,''
Nucl.\ Phys.\ B {\bf 375}, 119 (1992)
[arXiv:hep-th/9109005].
}

\lref\SenZF{
A.~Sen,
``Moduli space of unstable D-branes on a circle of critical radius,''
arXiv:hep-th/0312003.
}

\lref\SenMV{
A.~Sen,
``Remarks on tachyon driven cosmology,''
arXiv:hep-th/0312153.
}

\lref\FotopoulosYT{
A.~Fotopoulos and A.~A.~Tseytlin,
``On open superstring partition function in inhomogeneous rolling tachyon
background,''
JHEP {\bf 0312}, 025 (2003)
[arXiv:hep-th/0310253].
}

\lref\StromingerFN{
A.~Strominger and T.~Takayanagi,
``Correlators in timelike bulk Liouville theory,''
arXiv:hep-th/0303221.
}

\lref\TseytlinTI{
A.~A.~Tseytlin,
``Vector Field Effective Action In The Open Superstring Theory,''
Nucl.\ Phys.\ B {\bf 276}, 391 (1986)
[Erratum-ibid.\ B {\bf 291}, 876 (1987)].
}

\lref\OkuyamaWM{
K.~Okuyama,
``Wess-Zumino term in tachyon effective action,''
JHEP {\bf 0305}, 005 (2003)
[arXiv:hep-th/0304108].
}

\lref\GarousiAI{
M.~R.~Garousi,
``Slowly varying tachyon and tachyon potential,''
JHEP {\bf 0305}, 058 (2003)
[arXiv:hep-th/0304145].
}

\lref\KimIN{
C.~j.~Kim, Y.~b.~Kim and C.~O.~Lee,
``Tachyon kinks,''
JHEP {\bf 0305}, 020 (2003)
[arXiv:hep-th/0304180].
}

\lref\BraxRS{
P.~Brax, J.~Mourad and D.~A.~Steer,
``Tachyon kinks on non BPS D-branes,''
Phys.\ Lett.\ B {\bf 575}, 115 (2003)
[arXiv:hep-th/0304197].
}

\lref\SenBC{
A.~Sen,
``Open and closed strings from unstable D-branes,''
Phys.\ Rev.\ D {\bf 68}, 106003 (2003)
[arXiv:hep-th/0305011].
}

\lref\KwonQN{
O.~K.~Kwon and P.~Yi,
``String fluid, tachyon matter, and domain walls,''
JHEP {\bf 0309}, 003 (2003)
[arXiv:hep-th/0305229].
}

\lref\KlusonRD{
J.~Kluson,
``Particle production on half S-brane,''
arXiv:hep-th/0306002.
}


\rightline{EFI-04-01}
\Title{
\rightline{hep-th/0401066}}
{\vbox{\centerline{Notes on Tachyon Effective Actions}
\vskip 10pt \centerline{and Veneziano Amplitudes}}}
\bigskip
\centerline{Vasilis Niarchos}
\bigskip
\centerline{{\it Enrico Fermi Inst. and Dept. of Physics,
University of Chicago}}
\centerline{\it 5640 S. Ellis Ave., Chicago, IL 60637-1433, USA}
\bigskip\bigskip\bigskip
\noindent
In a previous paper (hep-th/0304045) it has been argued that
tachyonic Dirac-Born-Infeld (DBI) actions can be obtained
from open string theory in a limit, which generalizes the usual
massless DBI limit. In the present note we review
this construction focusing on a key property of the proposed
tachyon effective actions: how they reproduce appropriate
Veneziano amplitudes in a suitably defined kinematical region.
Possible extensions and interesting open problems are
briefly discussed.
\vfill

\Date{01/04}


\newsec{Introduction}

The appearance of a tachyon in the perturbative spectrum
of string theory (open or closed) signals the presence of an
instability that drives the theory away from the unstable
vacuum. During this time-dependent process the
tachyon mode grows exponentially and couples, in general,
to all the other modes of the string. The spacetime dynamics of this
process is captured by a nontrivial string field theory action,
whose formulation and analysis is a complicated problem. In
this note we argue, following \KutasovER, that
in certain limits in the configuration space of string theory
the tachyon decouples from the other massive stringy modes.
In such cases the spacetime dynamics of the tachyon is expected to be captured
by a much simpler effective action, whose formulation is the main
subject of this paper. This effective action is useful for
various purposes. It contributes to a further understanding
of the time-dependent dynamics of tachyon condensation in string theory
and provides a useful tool for interesting applications in more realistic setups
including recent discussions in cosmology (see, for example, \SenMV\ and
references therein).

To motivate our discussion, let us recall a more familiar case:
the massless Dirac-Born-Infeld (DBI)
effective action in open string theory \TseytlinDJ.
For our purposes it is enough to consider the ten-dimensional Born-Infeld
Lagrangian
\eqn\bi{
\LL_{\rm BI}=\sqrt{-\det(\eta_{\mu \nu}+2\pi \alpha' F_{\mu \nu})}
~, }
describing the dynamics of a $U(1)$ gauge field $A_{\mu}$ on
a single D-brane. Some of the key properties of this Lagrangian in the bosonic case
are the following:
\item{(1)} It has a surface of exact solutions parameterized by constant
$F_{\mu \nu}$ profiles. These are also exact solutions of the full open string
equations of motion, i.e.\ the massless vertex operator $F_{\mu \nu} X^{\mu} \p X^{\nu}$
is a true modulus of the theory.
\item{(2)} As an effective Lagrangian \bi\ is exact for arbitrary constant values of
$F_{\mu \nu}$ (the only restriction coming from a critical upper bound on the value
of the electric field).
\item{(3)} In string theory the general scattering amplitude of $n$ gauge bosons with
momenta $p_j$ and polarizations $\zeta_{\mu_j}$ (respecting the standard string
BRST conditions) takes the form
\eqn\gaubos{\eqalign{
&\II_n(\{\zeta_{\mu_j}\}, \{p_j\}) =
\int_{-\infty}^{\infty} \prod_{i=1}^{n-3} dy_i ~
\prod_{j=1}^n \zeta_{\mu_j}
\langle c \p X^{\mu_1}e^{ip_1 X} (0)
c\p X^{\mu_2}e^{ip_2 X} (1) \times
\cr
&\times c \p X^{\mu_3}e^{ip_3 X} (\infty)
\p X^{\mu_4}e^{ip_4 X} (y_1) \cdots \rangle
~ + ~{\rm inequivalent ~ orderings}
.}}
In this expression $c$ denotes the usual bosonic string ghost. To leading order
in the external momenta $p_i$ the $n$-point function \gaubos\ scales like
(momentum)$^n$. The same scattering amplitudes
can be computed from the Born-Infeld Lagrangian. They receive two
types of contributions: one comes from the one-particle-irreducible (1PI) diagrams
at order $n$ and the other from reducible exchange diagrams involving lower order irreducible
vertices connected by propagators. To leading order in momentum
these $n$-point functions again scale like (momentum)$^n$.
To order (momentum)$^n$, the string theory amplitudes \gaubos\ agree with
those obtained from the Born-Infeld action. An explicit check of this claim
up to 4th order can be found in \TseytlinTI.

\noindent
In fact, this is one way to determine the form of the Born-Infeld action.
We can write down the most general gauge-invariant local Lagrangian containing
all possible independent invariants of increasing dimensions and the properties
($1$)-($3$) will fix the form of this Lagrangian up to field redefinitions \TseytlinTI.

An alternative way to summarize the above set of properties is the following.
In the $\sigma$-model approach to string theory
(see, for example, \refs{\TseytlinRR,\TseytlinMT})
the configuration space of the theory can be viewed
as the space of (in general non-conformal) two-dimensional
worldsheet quantum field theories. The constant $F_{\mu \nu}$
profiles parameterize a surface of fixed points of the worldsheet
renormalization group (RG) and the Born-Infeld effective action governs
the dynamics of small (low-energy) fluctuations away from
this surface of fixed points.

Is it possible to find a suitable extension of this construction that incorporates
the dynamics of the open string tachyon? The massless DBI example
suggests the following course of action. First, we have to find a line of fixed
points of the worldsheet RG, i.e.\ tachyon profiles that are exactly
marginal. For example, we may consider the homogeneous rolling
tachyon solution of Sen \SenNU\ in open superstring theory\foot{In
this paper we discuss only the case of unstable non-BPS
D-branes in type II superstring theory. For relevant details
on the bosonic case see \refs{\KutasovER,\SmedbackUR}.}
\eqn\rolling{
T=T_+ e^{\frac{1}{\sqrt 2} x^0} +
T_- e^{-\frac{1}{\sqrt 2} x^0}
~ .}
This is parameterized by the arbitrary constants $T_{\pm}$ \foot{Strictly speaking,
inequivalent solutions are parameterized by only one parameter. This is
due to the freedom of shifting $x^0$ by an arbitrary constant.}
and is known to be an $exact$ solution of the full open
string equations of motion in the Euclidean signature
\refs{\CallanMW,\CallanUB,\PolchinskiMY}. The same is
believed to be true also in the Minkowki theory after the analytic
continuation $x^0 \rightarrow i x^0$.
The precise problem we want to solve is the following.
For starters, consider type II superstring theory in flat
space, i.e.\ in the absence of a rolling tachyon condensate.
We can write the following open string tachyon vertex operators
\eqn\minpic{
T^{(\pm)}_{\vec k}=e^{w_{\pm}x^0 + i\vec k \cdot \vec x}
~ .}
These vertex operators are in the ($-1$)-picture and the on-shell
condition reads\foot{In our conventions $\alpha'=1$.}
\eqn\disp{
w_{\pm}^2+\vec k^2=\frac{1}{2}
~ ,}
with the $\pm$ indices denoting the two opposite sign
solutions of this quadratic equation.
$\vec k =0$ for the rolling tachyon vertex operators
$e^{\pm \frac{1}{\sqrt 2} x^0}$ and small fluctuations around
them are given by small $spatial$ momenta and frequencies
\eqn\ww{
w_{\pm}=\pm \frac{1}{\sqrt 2} (1-\vec k^2)+O(\vec k^4)
.}
We want to study the leading order behaviour of the $n$-point Veneziano
amplitudes
\eqn\ven{
\AA_n = \langle T^{(\pm)}_{\vec k_1} T^{(\pm)}_{\vec k_2}
T^{(\pm)}_{\vec k_3} \cdots
T^{(\pm)}_{\vec k_n} \rangle
}
in the limit of small $spatial$ momentum. In the process of taking
this limit we have broken the explicit 10-dimensional Poincare
invariance, but this breaking is obviously spontaneous. The
choice of the time direction is arbitrary and the final expression of
the amplitudes $\AA_n$ will be Poincare invariant. In section 2 we find the following
structure. The amplitudes $\AA_n$ vanish automatically for $n$ odd and for $n$ even
they are certain Lorentz invariant functions of the
external momenta, which scale like (momentum)$^2$ to leading
order in spatial momenta. These quadratic functions can be split into two distinct pieces.
The first one is a quadratic polynomial $P_n$ of appropriate Mandelstam
variables. The second is a rational expression of the spatial momenta - let us call it
$W_n$ - that incorporates the complicated pole structure of
$\AA_n$ arising from intermediate on-shell tachyons. There are no massless
or higher mass stringy state poles. The higher mass stringy states are automatically
decoupled in the special kinematics of small spatial momentum. The form of $P_n$ is
completely fixed by symmetry considerations and it has $one$ free coefficient
at each order $n$. $W_n$ arises as a sum of lower order exchange diagrams and
therefore is fixed by induction.\foot{Compare this situation with the
corresponding one in the usual massless DBI case described above.
In that case the $n$-point functions scale like (momentum)$^n$ to leading order in momenta and
receive polynomial contributions from 1PI diagrams and a complicated pole
structure from exchange diagrams. In the tachyon case, the $n$-point amplitudes
scale like (spatial momentum)$^2$ to leading order in spatial momenta and receive a similar
type of contributions summarized in the functions $P_n$, $W_n$ above.}

We would like to obtain an action with the following well-defined properties:
\item{($a$)} The equations of motion of this action admit the rolling tachyon
solutions \rolling\ as exact solutions for any constants $T_{\pm}$.
\item{($b$)} The field theory amplitudes \ven\ obtained from this action
reproduce the corresponding string theory Veneziano amplitudes to the leading
quadratic order in spatial momenta, i.e.\ they have the leading order
structure described in the previous paragraph.

\noindent
If this action exists it will not be unique. String theory amplitudes are,
by definition, on-shell and there are well-known ambiguities in trying to
determine an effective Lagrangian purely from on-shell data. One has the freedom
to change its form performing field redefinitions
$T \rightarrow f(T,\p_{\mu}T,\p_{\mu}\p_{\nu}T,...)$ and/or adding couplings
that are proportional to the equations of motion. In section 3 these
ambiguities will be fixed in a very specific way. We consider
an effective Lagrangian of the form
\eqn\bbb{
\LL=\LL(T,\p_{\mu} T)
}
and demonstrate that this (single-derivative) ansatz is capable of satisfying
both requirements ($a$) and ($b$). Imposing condition ($a$)
fixes the form of the Lagrangian \bbb\ uniquely up to $one$ free coefficient
at each order in $T$. Let us call this coefficient $a_n$. Then, we compute
the leading order form of the amplitudes \ven\ in field theory and we find that they have the same
quadratic structure as in string theory. In particular, they contain $one$ free
parameter at each order $n$ and this parameter, which is a simple function of $a_n$,
is fixed by imposing requirement ($b$). This procedure determines $\LL$ completely up
to field redefinitions. As usual, different and more complicated actions with the
same properties can be obtained in different schemes.

So far we have considered string theory in flat space.
We would like to consider the extension of the above analysis
in the presence of non-vanishing tachyon condensates.
In section 4 we argue that if we can treat the theory perturbatively in $T_+$,
the same action will continue to reproduce the
leading quadratic terms of the perturbed string theory scattering amplitudes
\eqn\venn{
\langle T^{(\pm)}_{\vec k_1} T^{(\pm)}_{\vec k_2} \cdots T^{(\pm)}_{\vec k_n}
\rangle_{T_+}
}
in the presence of a half-brane rolling tachyon solution ($T_+$ arbitrary and
$T_-=0$ in \rolling). This extends the validity of this action in
the presence of a non-vanishing $T_+$-condensate. When we try to extend
the same reasoning in the full-brane case (both $T_+$ and $T_-$ non-vanishing in \rolling), we
encounter several complications and the action \bbb\ breaks down.
The properties of an alternative effective action are briefly discussed.

We would like to point out that the tachyon effective action obtained in the
above manner is a special case of a more general construction with the
following key characteristics. Around a line (or surface) of fixed points
in the configuration space of string theory small fluctuations of the
tachyon and/or massless modes decouple from the other massive modes
of the string. This decoupling limit generalizes the low-energy limit of the
massless case in an interesting new way, in which the excited fields fluctuate
infinitesimally around a line of classical solutions. In this note we
argue in favor of an effective action that describes correctly the interactions
of these small fluctuations. We do so in a special
setup in open string theory. Possible extensions to other cases and interesting open
problems are suggested in section 5.

\newsec{String theory Veneziano amplitudes with vanishing tachyon condensates}

To calculate the Veneziano amplitudes $\AA_n$ in \ven\ it is convenient to Wick rotate
$x^0 \rightarrow i x^0$ and consider the vertex operators \minpic\
in Euclidean signature. Then, in the vanishing spatial-momentum
limit the correlation functions \ven\ involve momentum modes, whose
momentum vectors are almost aligned with a particular, arbitrarily
chosen, axis in Euclidean space. We perform the computation of $\AA_n$
in this special kinematical region in Euclidean signature and at the end
we continue back to Minkowski space. At the current level of
understanding of string theory this is the standard way to compute
open string scattering amplitudes.

Odd-point functions will vanish automatically by momentum conservation.
On the other hand, non-vanishing $2n$-point functions involving the
positive frequency momenta $\{ k_1,...,k_n \}$ and the negative frequency
momenta $\{p_1,...,p_n\}$ have to satisfy the momentum conservation
constraints\foot{Henceforth this will be standard notation. Momenta
with a positive frequency will be denoted by $k_i$ and momenta with a
negative frequency will be denoted by $p_i$.}
\eqn\mom{\eqalign{
&\sum_{i=1}^n(\vec k_i +\vec p_i)=0 ~,
\cr
&\sum_{i=1}^n \vec k^2_i = \sum_{i=1}^n \vec p^2_i
~ .}}
They can be computed on the upper-half plane in the usual manner by
fixing the position of three vertex operators and integrating over the rest:
\eqn\even{\eqalign{
\AA_{2n}(\{\vec k_i\},\{\vec p_j\})&=g_o^{2n-2} \CC
\int_{-\infty}^{\infty} \prod_{i=1}^{2n-3} dy_i
\langle T^{(+)}_{\vec k_1}(0) T^{(+)}_{\vec k_2}(1)
\TTT^{(+)}_{\vec k_3}(\infty) \times
\cr
&\times \TTT^{(+)}_{\vec k_4}(y_1) \cdots
\TTT^{(+)}_{\vec k_n}(y_{n-3}) \TTT^{(-)}_{\vec p_1}(y_{n-2})
\cdots \TTT^{(-)}_{\vec p_n}(y_{2n-3})
\rangle + (k_2 \leftrightarrow k_3)
~ .}}
As usual, the total picture
number in superstring amplitudes must be equal to $-2$. In \even\
we have inserted two vertex operators in the ($-1$)-picture and
the rest in the 0-picture. In this paper we reserve the notation
$\TTT_{\vec k}^{(\pm)}$ for the 0-picture vertex operators
\eqn\zeropic{
\TTT^{(\pm)}_{\vec k}=i (\vec k \cdot \vec \psi - w_{\pm} \psi^0)
e^{i w_{\pm}x^0 + i\vec k \cdot \vec x}
~ .}
$g_o$ is the open string coupling and $\CC$ a universal
constant that can be determined by unitarity. For example, inspecting
the way that a 4-point function factorizes into the product of two 3-point
functions near a massless pole we obtain $\CC=1$
\PolchinskiRR.

The leading order form of these amplitudes can be reduced considerably
with a few simple observations. First of all, when we set the spatial
momenta $\{\vec k_i \}$, $\{ \vec p_j \}$ to zero, the amplitudes vanish
as a simple manifestation of the fact that the rolling tachyon
vertex operators $e^{\pm \frac{i}{\sqrt 2} x^0}$ are exactly marginal.
This is a special case of a more general statement. Correlation functions
of true moduli of string theory are identically zero. An interesting exception
to this rule, relevant for the massless DBI case, is the following.
Consider the $n$-point vector amplitudes \gaubos\ in bosonic string theory
(the situation in fermionic string theory is similar).
The first non-trivial term in the momentum expansion of the gauge boson
vertex operators $\VV_{\zeta,p} = \zeta_{\mu} \p X^{\mu} e^{i p \cdot X}$ has
the form $\zeta_{\mu} p_{\nu} X^{\nu} \p X^{\mu}$ and the amplitude
\eqn\zegabo{\eqalign{
&\II_n^{(0)}(\{\zeta_{\mu_j}\}, \{p_j\}) =
\int_{-\infty}^{\infty} \prod_{i=1}^{n-3} dy_i ~
\prod_{j=1}^n \zeta_{\mu_j}
\langle c p_{\nu_1}X^{\nu_1}\p X^{\mu_1} (0)
c p_{\nu_2} X^{\nu_2} \p X^{\mu_2}(1) \times
\cr
&\times c p_{\nu_3}X^{\nu_3} \p X^{\mu_3} (\infty)
p_{\nu_4}X^{\nu_4} \p X^{\mu_4}(y_1) \cdots \rangle
~ + ~{\rm inequivalent ~ orderings}
}}
gives the leading contribution to the full vector amplitude \gaubos\ in the
limit of small momenta. Indeed, this is one way to see that the leading term in \gaubos\ scales
like (momentum)$^n$ as we claimed above. The symmetric part of the vertex operator
$\zeta_{\mu} p_{\nu} X^{\nu} \p X^{\mu}$ is a trivial total derivative and the
antisymmetric part has the form $f_{\mu \nu} X^{\nu} \p X^{\mu}$, with
$f_{\mu \nu}$ a constant antisymmetric tensor. This is the familiar coupling
of a constant electromagnetic field on the boundary of the disk. It is a true modulus
of open string theory and the above general statement implies that
the corresponding amplitudes \zegabo\ have to vanish. An explicit calculation,
however, demonstrates that this conclusion is incorrect. As we
might expect also from the massless DBI effective action, the amplitudes
\zegabo\ are non-vanishing. The reason for this, perhaps unexpected, result is
the fact the vertex operators $\zeta_{\mu} p_{\nu} X^{\nu} \p X^{\mu}$ are
not well-defined CFT vertex operators and the computation of the corresponding
amplitudes is subtle.

In the case of the tachyon no subtleties of the above type appear and
the leading order form of the amplitudes \even\ is quadratic. This
quadratic dependence can be written as a function of the form
\eqn\rough{
\AA_{2n}(\{ \vec k_i, \vec p_j \})= g_o^{2n-2} \big ( P_{2n}
(\vec k_i,\vec p_j) + W_{2n}(\vec k_i, \vec p_j) \big )
~ .}
$P_{2n}(\vec k_i,\vec p_j)$ is a second order polynomial of the
spatial momenta and $W_{2n}(\vec k_i, \vec p_j)$ a second order rational
expression of the momenta with a sum of terms, each scaling like
(momentum)$^{2s+2}$/(momentum)$^{2s}$ (for appropriate term-dependent
positive integers $s$). The denominators of these terms are products of propagators
associated to intermediate nearly on-shell tachyons and the numerators
are determined uniquely as appropriate residues. A particular example of this
structure will be given below at 6th order. The higher order pole structure will
also be discussed.

The structure of the polynomial $P_{2n}$, on the other hand, is particularly
simple. It is straightforward to determine its exact form at any order
$2n$ up to a constant multiplicative factor by invoking the symmetry properties
of the amplitudes $\AA_{2n}$. This can be done most conveniently with
the use of the Mandelstam variables
\eqn\mandel{\eqalign{
S_{ij} &= (k_i +k_j)^2 \sim 2 - (\vec k_i -\vec k_j)^2 = 2 - s_{ij} ~,
\cr
T_{ij} &=(p_i+p_j)^2 \sim 2-(\vec p_i - \vec p_j)^2 =2- t_{ij} ~,
~ ~ i, j =1,...,n,
\cr
U_{ii'} &=(k_i+p_{i'})^2 \sim (\vec k_i +\vec p_{i'} )^2 = u_{ii'} ~,
~ ~ i, i'=1,...,n
~ ,}}
of which only a subset are linearly independent variables.
We have introduced the small letter symbols $s_{ij}$, $t_{ij}$ and
$u_{ii'}$, to denote particular quadratic combinations of the spatial
momenta, which will be referred to as the ``${\it  spatial}$
Mandelstam variables''. As we send $\{\vec k_i\}$, $\{\vec p_j\}$
to zero, the Mandelstam variables take the limiting values:
\eqn\limit{
S_{ij},T_{ij} \rightarrow 2, ~ ~ U_{ii'}=u_{ii'} \rightarrow 0,
~ ~ s_{ij}, t_{ij} \rightarrow 0
~ .}
With the use of these variables,
$P_{2n}$ can be cast into the Euclidean invariant form
\eqn\euclid{
P_{2n} =D+ \sum_{i,j} (A_{ij} S_{ij}+B_{ij} T_{ij}) +
\sum_{i,i'} C_{ii'} U_{ii'}
~ .}
$A_{ij}, B_{ij},C_{ii'}$ and $D$ are appropriate constants,
which can be constrained further by the
following symmetries of the amplitudes \even. First, there is a
symmetry under the exchange of the $\{k_i\}$ momenta among
themselves and similarly under the exchange of the $\{p_j\}$ momenta
among themselves. This property equates all the coefficients $A_{ij},
B_{ij}$ and $C_{ii'}$ to the constants $A, B$ and $C$ respectively.
Moreover, the  amplitudes \even\ are invariant under the transformation
$(x^0,\vec x)\rightarrow (-x^0,-\vec x)$, which exchanges the $\{k_i\}$
momenta with the $\{p_j\}$ momenta. This symmetry sets $A=B$ and
leads to the form
\eqn\euclidtwo{
P_{2n}=D + A\sum_{i,j}  (S_{ij}+T_{ij}) + C \sum_{i,i'} U_{ii'}
~ .}
One can easily check the identity
\eqn\stid{
\sum_{i,j} (S_{ij}+T_{ij}) = 4n^2 -2 \sum_{i,i'} U_{ii'}
~ ,}
which allows for a further simplification
\eqn\euclidthree{
P_{2n}=(D-4n^2 A) + (C-2 A) \sum_{i,i'} U_{ii'}
~ .}
As we mentioned above, the zeroth order term vanishes.
This sets $D=4n^2 A$ and after denoting the constant $C-2A$ by
$C_{2n}$ we are left with the unique (up to a multiplicative constant)
expression
\eqn\ptwo{
P_{2n}=C_{2n} \sum_{i,i'} U_{ii'} = C_{2n} \sum_{i,i'} u_{ii'}
~ .}

We proceed to verify and elucidate the general structure of eqs.\ \rough,
\ptwo\ with more explicit calculations - first at the lower 4th and 6th
orders and then inductively at the higher ones.

\subsec{4-point Veneziano amplitudes}

Setting $n=2$ in the general amplitude \even\ gives
\eqn\afour{\eqalign{
\AA_4&= g_o^{2} \int_{-\infty}^{\infty} dy
\langle e^{ik_1 \cdot x}(0) e^{ik_2 \cdot x}(1) (p_1 \cdot \psi)
e^{ip_1 \cdot x} (\infty) (p_2 \cdot \psi)e^{ip_2 \cdot x}(y) \rangle
+ (k_2 \leftrightarrow p_1) =
\cr
&= g_o^{2} (p_1 \cdot p_2) \int^{\infty}_{-\infty} dy
|y|^{2k_1\cdot p_2} |1-y|^{2k_2 \cdot p_2}
+ (k_2 \leftrightarrow p_1)
~ .}}
Further manipulation of this expression produces the
following function of the spatial Mandelstam variables
\eqn\aff{\eqalign{
\AA_4= g_o^{2} \big [ &(1-s_{12})B(u_{12'},u_{11'})
+(-1+u_{11'})B(u_{12'},2-s_{12})+
\cr
&+(-1+u_{12'})B(2-s_{12}, u_{11'}) \big ]
~ .}}
In standard notation, $B(s,t)$ denotes the Beta-function
\eqn\betaf{
B(s,t)=\int_0^1 dy ~ y^{-1+s} (1-y)^{-1+t} = \frac{\Gamma(u)
\Gamma(t)}{\Gamma(s+t)}
.}
\aff\ can be expanded to quadratic order in spatial momenta
(i.e.\ linear order in the spatial Mandelstam variables) using the
$\Gamma$-function expansion
\eqn\game{
\Gamma(\epsilon) \sim_{\epsilon \rightarrow 0} \frac{1}{\epsilon}
- \gamma + \frac{1}{2}(\gamma^2 + \zeta(2))\epsilon
~ ,}
where $\gamma$ denotes the Euler-Mascheroni constant and
$\zeta(2)=\frac{\pi^2}{6}$. At the end of this calculation we find
the leading order expression
\eqn\af{
\AA_4=-g_o^{2} \frac{\pi^2}{4}\sum_{ii'} u_{ii'}
~ ,}
which is precisely of the general form \rough, \ptwo, with
$C_4=-\frac{\pi^2}{4}$ and $W_4=0$. Notice the expected
vanishing of the 4-point amplitude at zero spatial momenta.

Another notable characteristic of the 4-point amplitude \af\ is the absence of any poles.
In general, string theory amplitudes possess a complicated pole structure
associated to the appearance of on-shell intermediate states.
An intermediate state can be a tachyon, a massless excitation
or a higher excited stringy state. In our case, the 4th order amplitude does not
exhibit any of these poles. It does not exhibit any tachyon poles, because of
the absence of a non-vanishing 3-point vertex among tachyons. Massless poles,
for example of the type $\frac{1}{u_{11'}}$ etc., cancel identically
after summing over all the channels
and massive stringy states only have finite contributions in the special kinematics
regime of this note. In higher orders, we expect to find a similar absence of massless
and higher mass poles. For the massless poles this is guaranteed only in the abelian
case, which is the case of interest in this paper. For the higher mass poles it is
a consequence of the special kinematics. The absence of these poles is
an important property of the present construction and is
equivalent to the statement of the introduction that small fluctuations around the
exactly marginal rolling tachyon direction decouple from any
other massive excitation of the open string. It is one of the key elements
that allows for the possibility of a consistent single-scalar tachyon
effective action.

\subsec{\it 6-point Veneziano amplitudes}

The 6th order Veneziano amplitude in string theory reads
\eqn\sixth{\eqalign{
\AA_6&=\langle T^{(+)}_{\vec k_1}T^{(+)}_{\vec k_2}
\TTT^{(+)}_{\vec k_3} \TTT^{(-)}_{\vec p_1}\TTT^{(-)}_{\vec p_2}
\TTT^{(-)}_{\vec p_3} \rangle =
\cr
&=\int^{\infty}_{-\infty} dx_1dx_2dx_3 \langle e^{ik_1 \cdot x}(0)
e^{ik_2 \cdot x}(1) (k_3 \cdot \psi)e^{ik_3 \cdot x}(\infty) \times
\cr
&\times (p_1 \cdot \psi) e^{ip_1 \cdot x}(x_1)(p_2 \cdot \psi)
e^{ip_2 \cdot x}(x_2) (p_3 \cdot \psi)e^{ip_3 \cdot x}(x_3) \rangle
+ (k_1 \leftrightarrow k_3)
~ .}}
By further evaluation we find
\eqn\sixx{\eqalign{
\AA_6&=(k_3 \cdot p_1)(p_2 \cdot p_3) \int^{\infty}_{-\infty}
\prod_i dx_i |x_1|^{2k_1\cdot p_1}|x_2|^{2k_1 \cdot p_2}
|x_3|^{2k_1\cdot p_3} \times
\cr
& \times |1-x_1|^{2k_2\cdot p_1}|1-x_2|^{2k_2\cdot p_2}
|1-x_3|^{2k_2\cdot p_3}|x_{12}|^{2p_1\cdot p_2}
|x_{13}|^{2p_1\cdot p_3}|x_{23}|^{-1+2p_2\cdot p_3}-
\cr
&-(k_3\cdot p_2)(p_1 \cdot p_3) \int^{\infty}_{-\infty} \prod_i dx_i
|x_1|^{2k_1\cdot p_1}|x_2|^{2k_1\cdot p_2}|x_3|^{2k_1\cdot p_3} \times
\cr
& \times |1-x_1|^{2k_2\cdot p_1}|1-x_2|^{2k_2\cdot p_2}
|1-x_3|^{2k_2\cdot p_3} |x_{12}|^{2p_1\cdot p_2}
|x_{13}|^{-1+2p_1\cdot p_3}|x_{23}|^{2p_2\cdot p_3}+
\cr
&+(k_3\cdot p_3)(p_1\cdot p_2) \int^{\infty}_{-\infty} \prod_i dx_i
|x_1|^{2k_1\cdot p_1}|x_2|^{2k_1\cdot p_2}|x_3|^{2k_1\cdot p_3} \times
\cr
& \times |1-x_1|^{2k_2\cdot p_1}|1-x_2|^{2k_2\cdot p_2}
|1-x_3|^{2k_2\cdot p_3}|x_{12}|^{-1+2p_1\cdot p_2}
|x_{13}|^{2p_1\cdot p_3}|x_{23}|^{2p_2\cdot p_3}+
(k_1 \leftrightarrow k_3)
}}
with a sum of six terms coming from every possible contraction of
the fermions. From the general considerations of the previous subsections
the leading order form of this amplitude is expected to become
\eqn\sixxform{
\AA_6=g_o^4\big ( C_6 \sum_{i,i'}u_{ii'} + W_6(\vec k_i, \vec p_j)
\big )
~ .}
Unfortunately, we are not aware of a closed expression for the above
integrals and this hinders the computation of the exact
value of the constant $C_6$. Nevertheless, it is quite straightforward
to determine the precise form of the rational function $W_6$. This
function involves a set of poles, which are associated to on-shell intermediate
tachyons. To be concrete, let us consider a particular tachyon pole, say the one
arising when an intermediate tachyon of momentum $k_1+k_2+p_1$ goes
on-shell. In this limit, the 6-point amplitude factorizes into the product of
two lower order 4-point amplitudes and the following expression is obtained
\eqn\sixfactor{
\AA_6(k_i,p_j) \sim \frac{\AA_4(k_1,k_2,p_1,-k_1-k_2-p_1)
\AA_4(-k_3-p_2-p_3,k_3,p_2,p_3)}{(k_1+k_2+p_1)^2-\frac{1}{2}}
~ .}
By using the leading order expression \af\ we find
\eqn\sss{
\AA_6(k_i,p_j) \sim \frac{\pi^4}{4}g_o^4\frac{s_{12}t_{23}}
{(k_1+k_2+p_1)^2-\frac{1}{2}}
~ .}
Summing over the nine different contributions of this type we obtain the
full dependence of the function $W_6$ on the spatial momenta $\{\vec k_i\}$,
$\{\vec p_j\}$.

\subsec{The pole structure of 2n-point Veneziano amplitudes}

Based on the above analysis (see eqs.\ \rough\ and \ptwo),
the general $2n$-point Veneziano amplitude \even\ can be written to
leading order as
\eqn\stwonvertex{
\AA_{2n}(\vec k_i,\vec p_j) = C_{2n} \sum_{i,i'} u_{ii'}
+W_{2n}(\vec k_i,\vec p_j)
~ .}
The $g_o$ dependence has been included implicitly into the definition
of the constant $C_{2n}$ and the function $W_{2n}$.
This function incorporates the complicated pole structure of
the full amplitude \even\ in the special kinematical regime of this note.
The general characteristics of $W_{2n}$ are the following.
There are many channels in \even, where poles appear. They arise from
regions of the moduli integrals in \even, where some number of the $y_i$
variables approach each other. For example, to analyze the limit where $2N$
of them approach 0, e.g.\ $y_1,y_2,...,y_N,y_{n-2},...,y_{n-N-1} \rightarrow 0$,
it is convenient to redefine
\eqn\ylim{
y_1=\epsilon, ~ y_2=\epsilon z_2, ...,~ y_N=\epsilon z_N,
~ y_{n-2}=\epsilon z_{n-2},..., ~y_{n-N-1}=\epsilon z_{n-N-1}
}
and consider the limit $\epsilon \rightarrow 0$ \DiFrancescoUD. The
residues of the resulting poles are related to lower order correlation functions of
intermediate states. Only poles associated to an intermediate tachyon are
interesting. Other types of poles do not appear for the reasons presented above.
For concreteness, let us consider the channel
\ylim, where an intermediate tachyon of momentum
$P=k_1+k_4+...+k_{N+3}+p_1+...+p_N$ goes on-shell. Near this pole
\even\ becomes
\eqn\factwon{
\AA_{2n} \sim \frac{\AA_{2N+2}(k_1,k_4,...,k_{N+3},p_1,...,p_N,P)
\AA_{2(n-N)}(-P,k_{N+4},...,k_n,p_{N+1},...,p_n)}{P^2-\frac{1}{2}}
~ ,}
which is a higher order analogue of \sixfactor. The form of the lower order
amplitudes to leading (quadratic) order is known by induction.
The full structure of $W_{2n}$ arises as a sum of similar contributions
over all possible channels. In this sum multiple poles scale as
(momentum)$^{2s+2}$/(momentum)$^{2s}$ for appropriate term-dependent
positive integers $s$.

\newsec{On-shell amplitudes in field theory and comparison with string theory}

In field theory we look for an effective tachyon Lagrangian with enough
parameters to reproduce the string theory scattering amplitudes \stwonvertex.
It has been proposed \KutasovER\ that the single derivative ansatz
\eqn\sing{
\LL=\LL(T,\p_{\mu} T)
~ ,}
symmetric under the parity transformation $T \rightarrow -T$, is a convenient
and consistent ansatz that has enough parameters to encode the full set of
string theory information to leading order. The even parity condition is an
immediate consequence of the standard $Z_2$ symmetry $\psi_{\mu}
\rightarrow -\psi_{\mu}$ of the fermionic string, under which the open string
tachyon is odd. The single derivative Lagrangian \sing\ fixes part of the field
redefinition ambiguity and leaves a residual freedom of taking
\eqn\tft{
T \rightarrow T f(T^2)
}
for an arbitrary function $f$.

In addition, assume that $\LL$ is analytic in $T$ around $T=0$. This assumption
is known to fail in the bosonic case \KutasovER, but no problems of this sort
are known to appear in the fermionic case. Then, we can expand $\LL$ in a
series of the form
\eqn\ltpt{
\LL=\sum_{n=0}^{\infty} \lambda^{2n-2} \LL_{2n}(T,\p_{\mu} T)
~ ,}
where
\eqn\ltwon{
\LL_{2n}=\sum_{l=0}^{n} a_l^{(n)} (\p_{\mu}T \p^{\mu}T)^l T^{2(n-l)}
}
and $\lambda$ is a constant related to the normalization of the tachyon $T$.
The problem of determining $\LL$ reduces into the problem of
determining the infinite set of constants $a_l^{(n)}$. This can be achieved
by imposing on $\LL$ the following two requirements that follow directly
from the discussion in the introduction:
\item{($A$)} The equations of motion of $\LL$ admit the generic rolling tachyon
profile \rolling\ as an exact solution.
\item{($B$)} The classical field theory $2n$-point amplitude reproduces exactly
the Veneziano $2n$-point amplitudes \even\ to leading
(quadratic) order in spatial momenta, i.e.\ it reproduces \stwonvertex\ for
any $n$.

The $2n$-th order equations of motion for \ltpt\ read \KutasovER\
\eqn\eqofmot{
\sum_{l=1}^nla_l^{(n)}\partial^\mu\left[
(\partial_\lambda T\partial^\lambda T)^{l-1}(\partial_\mu T)
T^{2(n-l)}\right]=
\sum_{l=0}^n(n-l)a_l^{(n)}T^{2(n-l)-1}(\partial_\mu T\partial^\mu T)^l
~ .}
By demanding that they admit
\eqn\rolrol{
T=T_+e^{\frac{i}{\sqrt 2}x^0}+T_- e^{-\frac{i}{\sqrt 2}x^0}
}
as an exact solution, we obtain at order $2n$ a set of $n$ linear equations
\eqn\recurse{
a_l^{(n)}=\frac{(n-1)!2^{l-1}}{(n-l)!l!(2l-1)}a_1^{(n)}
~ }
for the $n+1$ couplings $a_l^{(n)}$, ($l=0,1,...,n$).
In this way, the first requirement reduces the problem of determining $\LL$
drastically. At each order $only$ $one$ coefficient remains to be determined and
this will be done by imposing condition ($B$). This requires an explicit calculation
of the tree level field theory amplitudes to which we now turn. We
begin with a few illustrating examples at 4th and 6th order.

\subsec{Field theory 4-point amplitudes}

At 4th order the field theory couplings are summarized by the following Lagrangian
\eqn\lfour{
\LL_4= \lambda^2 \big (a_0^{(2)}T^4 + a_1^{(2)} (\p_{\mu} T \p^{\mu} T) T^2
+ a_2^{(2)} (\p_{\mu} T \p^{\mu} T)^2
\big )
.}
Translating this expression into momentum space, we obtain the 4-point function
(analogue of the string amplitude \afour)
\eqn\fafour{\eqalign{
\AA_4 &= \vertfour\ = \langle T^{(+)}_{\vec k_1} T^{(+)}_{\vec k_2}
T^{(-)}_{\vec p_1} T^{(-)}_{\vec p_2} \rangle =
\cr
&= \lambda^2 \big \{6a_0^{(2)}-a_1^{(2)}[(k_1 \cdot k_2)+(k_1 \cdot p_1)
+(k_1 \cdot p_2)+(k_2 \cdot p_1)+(k_2 \cdot p_2)+(p_1 \cdot p_2)]+
\cr
&+2a_2^{(2)}[(k_1 \cdot k_2)(p_1 \cdot p_2)+(k_1 \cdot p_1)(k_2 \cdot p_2)
+(k_1 \cdot p_2)(k_2 \cdot p_1)] \big \}
~ .}}
All momenta are on-shell and everything is considered in Euclidean signature
(the inner product between two vectors reads $k \cdot p = k^0 p^0
+\vec k \cdot \vec p$).
The expansion of this amplitude to quadratic order in spatial momenta gives
\eqn\faexp{
\AA_4=\lambda^2 \big (6a_0^{(2)}+a_1^{(2)}+\frac{3}{2} a_2^{(2)} \big ) -
\lambda^2 a_2^{(2)} \sum_{i,i'=1}^2 u_{ii'}
~ .}
This expression has the same form as its string theory counterpart \af\ and
a direct comparison between them yields the following two conditions on the
$\LL_4$ coupling constants:
\eqn\ccfour{\eqalign{
&6a_0^{(2)}+a_1^{(2)}+\frac{3}{2} a_2^{(2)} = 0,
\cr
&a_2^{(2)}=\lambda^{-2}g_o^2 \frac{\pi^2}{4}
~ .}}

The first condition comes from the vanishing of the constant term in the
string theory amplitude \afour, which follows from the fact that the rolling
tachyon profile is exactly marginal. In field theory, the analogous requirement
(condition ($A$) above) gives at the $2n$-th order $n$ constraints, which lead to
the recursion relation \recurse. For $n=2$ one of these constraints should
be coincident with the first equation in \ccfour. We can see this,
generalized to any order, in the following way.

Substitute the rolling tachyon profile \rolrol\ into the equations of
motion \eqofmot\ and set the coefficients of the $n$
exponentials $e^{i\frac{m}{\sqrt 2}x^0}$ (for $m=1,3,...,2n-1$) to
zero.\foot{There is an extra set of equations originating from the vanishing of the
coefficients of the exponentials $e^{-i\frac{m}{\sqrt 2}x^0}$ ($m=1,3,...,2n-1$).
This set is identical to that of the positive frequency exponentials and will not be
discussed independently.} These equations effectively set to zero all tree-level
$2n$-point vertices of the form
\eqn\eomvert{
\pmone ~ , \pmtwo ~ , ~ ... ~ , \pmn
,}
with the last rightmost vertex having $n$ positive incoming frequencies and $n-1$
negative incoming frequencies. All these vertices (in total $n$ in number) are
evaluated at zero spatial momenta and they provide a set of equations
equivalent to the recursion relations \recurse. The momentum of the outgoing particle is
determined by momentum conservation and can be on-shell only for the last rightmost
diagram, where the positive incoming frequencies almost cancel the negative incoming
frequencies. This latter condition is the only one appearing in the evaluation of on-shell
amplitudes.

To illustrate this point more explicitly let us consider in detail the example of the 4-point
vertices. For $n=2$ we find two vertices, which upon
evaluation give
\eqn\feomvert{
\ppfone ~ = \lambda^2 \big (2a_0^{(2)}+a_1^{(2)}-\frac{3}{2} a_2^{(2)}\big ),
}
\eqn\ffeomvert{
\ppftwo ~ = \lambda^2 \big (6a_0^{(2)}+a_1^{(2)}+\frac{3}{2} a_2^{(2)} \big )
.}
The second condition is the only one appearing in the on-shell 4-point amplitude
\faexp. The vanishing of the first vertex \feomvert\ is an off-shell statement, which will be
imposed here as part of condition ($A$) of the previous subsection. We see that
conditions ($A$) and ($B$) are not completely independent, but they have a small overlap.

At non-zero spatial momenta comparison with the on-shell string theory amplitude
provides one more equation, the second one in \ccfour. In higher orders this extra
equation, together with the $n$ ones coming from condition ($A$), is enough to
determine the exact form of the effective Lagrangian order by order. In 4th order,
we simply get a relation between the coefficient $a_2^{(2)}$, the normalization factor
of the tachyon $\lambda^2$ and the open string coupling $g_o$.

\subsec{Field theory 6-point amplitudes}

The field theory 6-point amplitude
\eqn\sixam{
\AA_6=\langle T^{(+)}_{\vec k_1} T^{(+)}_{\vec k_2} T^{(+)}_{\vec k_3}
T^{(-)}_{\vec p_1} T^{(-)}_{\vec p_2} T^{(-)}_{\vec p_3} \rangle
}
receives several contributions. One comes directly from the 6th order part of
the effective tachyon Lagrangian
\eqn\sixlag{
\LL_6=\lambda^4 \sum_{l=0}^3 a_l^{(3)} (\p_{\mu}T \p^{\mu} T)^l T^{6-2l}
}
and the rest come from exchange diagrams involving 4th order vertices. In
diagrammatic form we have
\eqn\sixdiag{
\AA_6=\blobsix\ = \vertsix\ +\plainsix\
~ .}
There are two types of exchange diagrams. The first one
\eqn\kkk{
\kkksix
}
involves 4-vertices of the form
\eqn\kkkvert{
\ppfone
}
and the propagator is always off-shell and regular even when the spatial momenta
vanish. The second type involves 4-vertices of the form
\eqn\kkpvert{
\ppftwo ~ ,
}
which become on-shell as we send the spatial momenta to zero and develop a
pole. A particular example includes the diagram
\eqn\kkp{
\kkpsix
~ .}
There are nine different choices of this type and each of them contributes
non-polynomial terms in 6th order, which scale like
(momentum)$^4$/(momentum)$^2$ to leading order in spatial momenta.
Analogous exchange diagrams appeared in string theory (see eq.\ \sixfactor).
A more detailed analysis provides the following information.

The 6th order vertex arising from \sixlag\ (here we demonstrate explicitly only
a few representative terms) reads
\eqn\qqq{\eqalign{
\lambda^{-4} \vertsix &=\frac{1}{20}\bigg[ 6! a_0^{(3)}- 2 \cdot 4! a_1^{(3)}
\big ( k_1 \cdot (k_2+k_3+p_1+p_2+p_3) +
\cr
& +k_2 \cdot (k_3+p_1+p_2+p_3)+
k_3 \cdot (p_1+p_2+p_3)+p_1 \cdot (p_2+p_3)+p_2 \cdot p_3 \big ) +
\cr
&+ 16 a_2^{(3)} \big ( (k_1 \cdot k_2) (k_3 \cdot (p_1+p_2+p_3))+ \cdots \big )-
\cr
&- a_3^{(3)} \big ( (k_1 \cdot k_2) (k_3 \cdot p_1) (p_2 \cdot p_3) +\cdots \big )
\bigg]
~ .}}
Simple symmetry considerations (similar to those presented for string theory
amplitudes above) reduce the form of this expression in the limit of small spatial
momenta to
\eqn\hhh{
\vertsix = t_6+s_6 \sum_{i,i'} u_{ii'}
~ .}
$t_6$ and $s_6$ are certain constants that can be determined by a straightforward
calculation
\eqn\tsconst{\eqalign{
t_6&=\frac{\lambda^4}{20} \big ( 6! a_0^{(3)}+3 \cdot 4! a_1^{(3)}+36 a_2^{(3)}
+\frac{6!}{8} a_3^{(3)} \big ),
\cr
s_6&=\lambda^4 \big (-2a_2^{(3)}+9a_3^{(3)} \big )
~ .}}

The regular exchange diagrams can be expanded to quadratic order in a similar
fashion. We find:
\eqn\uuuquad{\eqalign{
&\kkksix = \frac{9}{4} \lambda^4 \big (2a_0^{(2)}+a_1^{(2)}
-\frac{3}{2}a_2^{(2)} \big )^2 +
\cr
&+\frac{3}{64} \lambda^4 \big ( 2a_0^{(2)}+a_1^{(2)}-\frac{3}{2} a_2^{(2)}
\big ) \big( 12 a_0^{(2)}-2a_1^{(2)}+31a_2^{(2)} \big ) \sum_{i,i'} u_{ii'}
~ .}}
Both terms in this expression are proportional to the linear combination
$2a_0^{(2)}+a_1^{(2)}-\frac{3}{2} a_2^{(2)}$ appearing in \feomvert.
This combination is set to zero by requirement ($A$) and the regular set of
exchange diagrams does not contribute in 6th order.

The second type of exchange diagrams involves a nearly on-shell pole. Let us
consider explicitly one of them, say the one appearing in \kkp.
To leading order in spatial momenta we find:
\eqn\xxxquad{\eqalign{
&\kkpsix = \frac{1}{-s_{12}+u_{11'}+u_{22'}}\bigg [ \big (6a_0^{(2)}+a_1^{(2)}
+\frac{3}{2} a_2^{(2)} \big )^2 +
\cr
&+(-s_{12}+u_{11'}+u_{22'})\big ( 6a_0^{(2)}+a_1^{(2)}
+\frac{3}{2} a_2^{(2)} \big )\big ( a_1^{(2)}+3a_2^{(2)} \big )
\cr
&+(-s_{12}+u_{11'}+u_{22'})^2\big ( \frac{1}{4}\big( a_1^{(2)}\big )^2
+a_1^{(2)}a_2^{(2)}+\frac{5}{2}\big( a_2^{(2)} \big )^2 \big ) +
\cr
&+(-s_{12}+u_{11'}+u_{22'})(s_{12}+t_{23})\big ( \big (a_2^{(2)}\big)^2
-a_1^{(2)}a_2^{(2)} \big )+
\cr
&+4\big (a_2^{(2)}\big)^2 s_{12}t_{23} \bigg ]
~ .}}
The first two terms are proportional to the linear combination
$6a_0^{(2)}+a_1^{(2)}+\frac{3}{2} a_2^{(2)}$,
which has been set to zero (see eq.\ \ccfour). The third and fourth
terms are non-vanishing polynomial quadratic contributions and the last
one is a pole contribution that scales like (momentum)$^4$/(momentum)$^2$.
Using eqs.\ \ccfour, \xxxquad\ and \sss, we can check that the
same non-polynomial term appears in string theory as part of the function $W_6$.

The full 6th order amplitude is obtained by summing over the individual contributions
\hhh, \xxxquad:
\eqn\sixtotal{\eqalign{
&\blobsix=\lambda^4 \bigg [\frac{1}{20}\big ( 6! a_0^{(3)}+3 \cdot 4! a_1^{(3)}
+36 a_2^{(3)}+\frac{6!}{8} a_3^{(3)}\big)+
\cr
&+\bigg( -2a_2^{(3)}+9a_3^{(3)} + \frac{1}{8}\big(a_1^{(2)}\big)^2-
\frac{13}{4}a_1^{(2)}a_2^{(2)}+\frac{25}{8}\big(a_2^{(2)}\big)^2 \bigg)
\sum_{i,i'}u_{ii'}+
\cr
&+4\big( a_2^{(2)} \big)^2 \bigg ( \frac{s_{12}t_{23}}{-s_{12}+u_{11'}
+u_{22'}}+\cdots \bigg ) \bigg ]
~ .}}
In the derivation of this expression we have enforced the 4th order
vanishing conditions \feomvert, \ffeomvert\ and by $\cdots$ in the last term
a summation over the remaining eight channels of the type \kkp\ is implied.
We can further check the vanishing of the constant term
$6! a_0^{(3)}+3 \cdot 4! a_1^{(3)}+36 a_2^{(3)}+\frac{6!}{8} a_3^{(3)}$
as a consequence of the recursion relation \recurse.

In conclusion, we see explicitly that the structure of the 6th order
tree-level amplitudes in field theory is precisely the same to
the one encountered in string theory above. The leading term in spatial momenta scales like
(momentum)$^2$ and breaks up into a quadratic polynomial piece and a rational
function (denoted by $W_6$ in \sixxform) coming from exchange diagrams.

\subsec{Field theory $2n$-point amplitudes}

The discussion of the 6th order amplitudes can be extended naturally to any
order. In the limit of vanishing spatial momenta the full $2n$-point amplitudes in field
theory take the form
\eqn\twonvertex{
\blobtwon =\tilde C_{2n} \sum_{i,i'} u_{ii'} +\tilde W_{2n}
(\vec k_i,\vec p_j)
~ .}
A possible constant term vanishes after imposing the conditions coming from
requirement ($A$) (see eq.\ \recurse) up to order $2n$. $\tilde C_{2n}$ is a
constant arising from the leading order expansion of two sets of diagrams.
The first (regular) set includes the irreducible $2n$-vertex
\eqn\twonnn{
\verttwon
}
and exchange diagrams, which are analogs of \kkk. The propagators
appearing in these diagrams are regular in the special kinematical regime of
this paper and yield a finite contribution. Their leading order momentum
dependence is a quadratic polynomial. The second set includes a series of
exchange diagrams with $k \neq 0$ near on-shell
propagators. Particular examples include the diagrams
\eqn\polediags{
\kkpkkp ~ , ~ ~ \kkpcube
~ }
with $k=1,2$ respectively.
These diagrams are higher order analogs of those appearing in \kkp.
The leading order momentum dependence of such diagrams can be summarized by a quadratic
polynomial, which contributes to the constant  $\tilde C_{2n}$, and a rational
function of the external momenta, which contributes to the function $\tilde W_{2n}$
in the same way as in string theory.
The poles appearing in this function are identical to the poles appearing
in the string theory function $W_{2n}$ and the residues are derived
in both cases as appropriate factorizations of the same lower order amplitudes.
This implies the relation
\eqn\wtw{
\tilde W_{2n}=W_{2n}
}
for any $n$. We verified this relation explicitly for $n=2,3$ above.

\subsec{Computation of the effective action}

We have demonstrated that in field theory the leading order
 amplitudes take the form
\eqn\field{
\AA_{2n}= \bigg( \sum_{l=1}^n d^{(n)}_l a_l^{(n)}
+ q^{(n)}_{\rm lower ~ order} \bigg)
\sum_{i,i'} u_{ii'} + W_{2n}(\vec k_i,\vec p_j)
~ ,}
with $d^{(n)}_l$ certain constants that arise
from the field theory vertex \twonnn\ (see appendix A) and
$q^{(n)}_{\rm lower ~ order}$ a constant that arises from
lower order exchange diagrams. In string theory the respective
expression reads
\eqn\ooo{
\AA_{2n}=C_{2n} \sum_{i,i'} u_{ii'}+W_{2n}(\vec k_i,\vec p_j)
~ ,}
for some constant $C_{2n}$. The same
function $W_{2n}$ appears on both sides and
the task of matching the two results amounts to matching the
coefficients of the $\sum_{i,i'} u_{ii'}$ term.
This gives an extra condition at the $2n$-th order in field theory, which
fixes the full set of $n+1$ couplings $a_l^{(n)}$ in \ltwon\ and
determines the form of the effective field theory action.

Indeed, the other $n$ conditions among the couplings $a_l^{(n)}$ arise from
condition ($A$) in the beginning of this section. They are summarized
in relation \recurse. One could use this relation to re-express the sum
$\sum_{l=1}^n d^{(n)}_l a_l^{(n)}$ appearing in \field\ in terms of
just one coefficient, say $a_1^{(n)}$ and the resulting expression
can be written as $d^{(n)} a_1^{(n)}$ for an appropriate constant
$d^{(n)}$. Then, matching equations \field\ and \ooo\ amounts to setting
\eqn\zzz{
d^{(n)}~ a_1^{(n)} + q^{(n)}_{\rm lower ~ order} = C_{2n}
~ .}
This determines the unknown coupling constant $a_1^{(n)}$ in terms of
$d^{(n)}, q^{(n)}_{\rm lower ~ order}$ and $C_{2n}$. Subtleties could
arise if, for some $n$, we happened to find $d^{(n)}=0$. In that case eq.\ \zzz\
would not be satisfied for any value of $a_1^{(n)}$, unless
$q^{(n)}_{\rm lower ~ order} = C_{2n}$. An explicit alternating series
expression for $d^{(n)}$ is given in appendix A. Using Mathematica, we
have checked that up to $n=15$ the absolute value of $d^{(n)}$ is a non-vanishing
monotonically increasing function of $n$. We expect a similar statement
to be true for all $n$.

In summary, the above discussion demonstrates the existence of a
tachyon effective action with a prescribed set of properties.
The form of this action can be fixed uniquely (modulo the usual
field redefinition ambiguities) so long as we know the precise
value of the constants $d^{(n)}, q^{(n)}_{\rm lower ~ order}$
and $C_{2n}$. This turns out to be a non-trivial technical problem
involving, in particular, the computation of a series of complicated
integrals (see, for example, \sixx).

\newsec{Effective actions in the presence of tachyon condensates}

The analysis of the previous sections established that up
to field redefinitions there exists a unique tachyon effective
action satisfying requirements ($A$) and ($B$) of section 3
in the absence of a tachyon condensate. Can we trust the
same action also in the presence of a non-zero tachyon
condensate? For starters, let us consider turning on a
non-zero constant $T_+$ in the rolling tachyon profile \rolrol,
while keeping $T_-$ zero. In the presence of this
half-brane rolling tachyon the spectrum of
scaling dimensions of the theory remains unmodified. If
the theory can be treated perturbatively in $T_+$,
the effective action of section 3 can still be used to
describe the leading terms of the string theory
scattering amplitudes \venn\ in the limit of small spatial momenta
and this extends its validity in the presence of a
non-vanishing $T_+$-condensate.

On a practical level, it would be useful to have more than a
statement about the existence of this action. It would be useful
to obtain the exact form of the Lagrangian \ltpt, \ltwon.
This is possible in the language of section 3, but it involves a
series of complicated calculations. In order to avoid these
technical difficulties, the analysis of \KutasovER\ proposed that the
exact form of the effective action \ltpt, \ltwon\ can be fixed
with an alternative calculation. This approach is based on the
evaluation of the ``unintegrated'' disc partition function in Minkowskian
signature. Let us briefly explain the basic features of this computation.
On general grounds, one expects that the on-shell spacetime action is
equal to the perturbed disc partition sum \refs{\TseytlinRR,\WittenQY
\TseytlinMT\GerasimovZP\KutasovQP\KutasovAQ
\MarinoQC-\NiarchosSI}. By stripping off the integral
over the zero-mode $x^0$, it is natural to conjecture the closely
related formula
\eqn\lzrel{
\LL_{\rm on-shell}(x^0)=Z'(x^0)
~ .}
$\LL_{\rm on-shell}(x^0)$ is the on-shell value of the effective
Lagrangian and $Z'(x^0)$ is the disc partition sum
\eqn\dipasu{
Z'(x^0)=\int [d{x'}^0] ~ e^{-\SS_{\rm bulk} - \int_{\p D} d\sigma T(x^0)}
}
where the disc path integral is performed only over the non-zero
modes ${x'}^0$; the zero modes remain unintegrated and
momentum conservation in the $x^0$ direction is not imposed.
In our case, both sides of \lzrel\ should be evaluated on the half-brane tachyon
 profile $T=T_+ e^{\frac{1}{\sqrt 2}x^0}$. The value of $Z'(x^0)$
for this profile has been computed in \LarsenWC:
\eqn\larsen{
Z'(x^0)=\frac{1}{1+\frac{1}{2}T_+^2 e^{\sqrt 2 x^0}}
~ .}
With this information one can show, by a direct application of
\lzrel, that the effective Lagrangian takes the form \KutasovER
\eqn\lkn{
\LL=-\frac{1}{1+\frac{1}{2}T^2}
\sqrt{1+\frac{1}{2}T^2+\p_{\mu}T\p^{\mu}T}
~ ,}
which transforms into the more familiar tachyon DBI
action\foot{As noted in \KutasovER\ the form of this Lagrangian
is quite different from the one found in boundary string field theory (BSFT)
\refs{\GerasimovZP,\KutasovQP,\KutasovAQ}. From the point
of the present discussion this is quite natural. The BSFT
action is valid for deeply off-shell tachyons, e.g.\ of the form
$T = a+u x^2$ in the bosonic case, while the tachyon DBI action
is valid for nearly on-shell configurations of the form \rolling.}
\refs{\SenMD\GarousiTR\BergshoeffDQ\KlusonIY\GibbonsHF
\LambertFA\SenNU\SenIN\SenAN\SugimotoFP\MinahanIF\IshidaFR
\OhtaAC\LambertHK\GibbonsTV\KimHE\KimQZ\LeblondDB\LambertZR\GarousiPV
\OkuyamaWM\GarousiAI\KimIN\BraxRS\SenBC\KwonQN-\KlusonRD}
\eqn\tdbi{
\LL=-\frac{1}{\cosh \frac{\tilde T}{\sqrt 2}}
\sqrt{1+\p_{\mu} \tilde T \p ^{\mu} \tilde T}
}
after the field redefinition
\eqn\fredef{
\frac{T}{\sqrt 2}=\sinh \frac{\tilde T}{\sqrt 2}
~ .}

This approach is similar to the usual derivation of
the massless Born-Infeld action. In that case, one is instructed to
compute the disc partition sum in the presence of a constant
$F_{\mu \nu}$ profile. It is natural to expect that this latter derivation of
the tachyon effective action is equivalent to the one described
in section 3 and that both of them lead to the same result.
It would be interesting to compute the constants $C_{2n}, \tilde C_{2n}$
in \stwonvertex\ and \twonvertex\ respectively and check this
claim explicitly. This would also provide a non-trivial check of the
assumption \lzrel.

More generally, we can ask whether the same effective
action is valid in the presence of other non-vanishing
tachyon condensates. For example, consider turning on
both $T_+$ and $T_-$ in \rolling. In that case several complications
appear. One drastic effect of the condensate is the modification
of the spectrum. Another related effect, most apparent in the
Minkowskian signature, is the absence of asymptotic flat
regions in spacetime. At very early ($x^0 \rightarrow -\infty$)
and very late ($x^0 \rightarrow +\infty$) times the system
is very far from its perturbative open string vacuum and
the meaning of S-matrix elements is far from obvious. In fact,
as $|T| \rightarrow \infty$ the theory approaches the
closed string vacuum, where no physical open string excitations
are believed to survive. These effects invalidate the action
of section 3 as a reliable tachyon effective action. Indeed, this
breakdown has been observed directly
in the computation of the stress-energy tensor
in refs.\ \refs{\SenIN, \LambertZR, \KutasovER}. Despite of
these problems, the action \tdbi\ continues to be valid at late times.
As $x^0 \rightarrow +\infty$ the exponential term
$T_- e^{-\frac{1}{\sqrt 2} x^0}$ can be ignored in the rolling tachyon profile
\rolling\ and the action \tdbi\ resurfaces at late times as a valid description
of the appropriate string theory dynamics.

It would be very interesting to see whether it is possible
to obtain another effective action in the presence of the full-brane
tachyon condensate. If such an action exists it will not
be Poincare (Euclidean) invariant. Poincare
invariance is explicitly broken by the full-brane condensate.\foot{The
half-brane condensate allows for a manifestly
Poincare invariant flat region in the far past and does not
break Poincare invariance.} The band
structure of the open string theory spectrum in the presence
of the full-brane tachyon \CallanUB\ suggests that the
desired action could be some kind of non-commutative
deformation - possibly a q-deformation - of the original tachyon DBI
action \tdbi. Whether this expectation is true remains to be seen.

\newsec{Conclusions and open problems}

What are the consequences of the present construction for
open string tachyon condensation? Building on previous work
\KutasovER, we argued that the dynamics of the open string
tachyon in the vicinity of the homogeneous half-brane solution
are captured correctly by the effective action \tdbi. The
validity of this effective action is not restricted only at late
times, where the gradients of the tachyon potential vanish
asymptotically. The action describes correctly the process
of tachyon condensation everywhere along the decay. By that
we mean precisely the following. As long as the initial conditions
are such that the tachyon never deviates considerably away
from the homogeneous half-brane profile (i.e.\ the tachyon
has the general inhomogeneous form
$T=T_+(x^{\mu})e^{\frac{1}{\sqrt 2}x^0}+T_-(x^{\mu})
e^{-\frac{1}{\sqrt 2} x^0}$, with
$|\p_{\mu}T(x^{\nu})|\ll 1, |T_-(x^{\nu})|\ll 1$)
the effective action \tdbi\ provides a valid description of the
dynamics and will not break down. This is also in nice agreement with some
features of general inhomogeneous solutions of this action discussed
in refs.\ \refs{\FelderSV,\BerkoozJE} (for more details see \KutasovER).

The usefulness of the action \tdbi\ is comparable to that of the
usual massless DBI action. Both actions capture reliably a
class of phenomena in the full open string theory that can be
associated to small deviations away from the corresponding exactly marginal
profiles. It is worth mentioning the following two prominent examples
in the tachyon case (again, for more details we refer the reader to \KutasovER):
\item{($i$)} The tachyon DBI action \tdbi\ reproduces the correct stress-energy
tensor in homogeneous tachyon decay for the half-brane
solution and the leading $T_-$ behaviour of the stress-energy tensor
for the full-brane solution.
\item{($ii$)} It contains solitonic solutions corresponding
to lower dimensional stable D-branes \SenTM. Small excitations of these
solitons are massless fields (similar to the gauge fields $A_{\mu}$
and scalars $Y^I$ on a stable D-brane) and one can check explicitly \SenTM\
that the effective action describing the dynamics of these modes is the massless
DBI action, exactly as anticipated from string theory.

We would like to conclude with a few interesting open problems.
One of them is the extension of the action \tdbi\ to
include the gauge field $A_{\mu}$ and the massless scalars $Y^I$
parameterizing the position of the D-brane. This is particularly
simple when $F_{\mu 0}=\p_0 Y^I=0$. In that case the rolling
tachyon solution \rolling\ is not modified by the expectation
values of the massless fields and the full action is very likely to
be given by the DBI generalization
\eqn\dbigen{
\LL=-\frac{1}{\cosh \frac{\tilde T}{2}}\sqrt{- \det G}
~ ,}
with
\eqn\gmn{
G_{\mu \nu} = \eta _{\mu \nu} +\p_{\mu} \tilde T
\p_{\nu} \tilde T + \p_{\mu} Y^I \p_{\nu} Y^I + F_{\mu \nu}
~ .}
In the presence of a nonvanishing electric field $F_{\mu 0}$ and
non-vanishing D-brane velocities $\p_0 Y^I$ the above extension
is less straightforward. Presumably one has to replace \rolling\
by a solution of the tachyon equations of motion in the open string
metric and proceed from there. For a recent discussion in favor
of the action \dbigen, \gmn\ see \SenZF.

The case of a non-zero electric field is also interesting for another
reason.\foot{I would like to thank D. Kutasov for suggesting this
interesting point.} A constant electric field couples on the boundary
of the disc with the coupling
\eqn\electric{
\int_{\p \Sigma} d\sigma ~ F_{i0} x^i \p x^0
~ .}
The Euclidean version of the $x^0$ CFT has an algebraic $SU(2)$
symmetry under which $i\p x^0$ rotates, for example, into
$e^{\frac{i}{\sqrt 2}x^0}+e^{-\frac{i}{\sqrt 2} x^0}$. This
suggests that \electric\ has the following equivalent form\foot{The
presence of the $x^i$ CFT should not affect this statement. We
could apply the $SU(2)$ symmetry of the $x^0$ CFT perturbatively
in $F_{0i}$.}
\eqn\inho{
\int_{\p \Sigma} d\sigma ~ F_{i0} x^i \big(
e^{\frac{i}{\sqrt 2}x^0}+e^{-\frac{i}{\sqrt 2} x^0} \big )
~ .}
This is a (linear) inhomogeneous version of the full-brane
rolling tachyon solution. A similar linear inhomogeneous version of
the half-brane rolling tachyon solution has been considered in a
recent paper \FotopoulosYT. It would be interesting to explore
the properties of open string theory in the presence of such condensates
and see to what extent the algebraic $SU(2)$ symmetry is a useful tool.

Finally, it would be very interesting to see if the general point of
view on effective actions reviewed here is useful for thinking about
analogous examples in closed string tachyon condensation. Progress
in this direction requires establishing the existence of an exact
closed string tachyon solution analogous to the rolling tachyon solution in
open string theory. For a previous suggestion that the perturbation
\eqn\clcl{
\delta \LL_{\rm ws}=\lambda e^{2x^0}
}
is exactly marginal see ref.\ \StromingerFN.

\vskip 1cm
\noindent{\bf Acknowledgements:}

I would like to thank David Kutasov and Arkady Tseytlin for many
stimulating discussions and for useful suggestions on the manuscript. Especially,
I would like to thank David Kutasov for earlier collaboration
and for his large contribution to this project, for his encouragement and
constant support. This work was supported in part by DOE grant
DE-FG02-90ER-40560.

\appendix{A}{Calculation of the field theory amplitude $\verttwon$}

In this appendix we compute the coefficient of the $\sum_{i,i'} u_{ii'}$ term
in the field theory amplitude $\verttwon$ at order $2n$. This computation can be
achieved compactly in the following way. Set
\eqn\tach{
T=\sum_{i=1}^n (e^{i k_i \cdot x} + e^{i p_i \cdot x} )
}
into the field theory Lagrangian
\eqn\qqq{
\LL_{2n}=\sum_{l=0}^{n} a_l^{(n)} \big ( \p_{\mu} T \p^{\mu}
T \big ) ^l T^{2(n-l)}
}
and pick up those terms that satisfy the momentum conservation constraints
$\sum_i (k_i +p_i)=0$. For simplicity, we have omitted a symmetry factor and
the normalization constant $\lambda^{2n-2}$. The resulting expression can be
expanded to quadratic order and, as we know on general grounds, the final result
will take the form
\eqn\eee{
t_{2n}+s_{2n} \sum_{ii'} u_{ii'}
~ .}
We would like to determine the coefficient $s_{2n}$. A convenient choice of
the external momenta simplifies the computation considerably. If we set
\eqn\choice{\eqalign{
&k_1=(\frac{1}{\sqrt 2}(1-\sigma^2), \sigma, 0,...),
~ k_2=k_3=...= k_n=(\frac{1}{\sqrt 2},0,0,...) ,
\cr
&p_1=(-\frac{1}{\sqrt 2}(1-\sigma^2), -\sigma, 0,...),
~ p_2=p_3=...= p_n=(-\frac{1}{\sqrt 2},0,0,...)
}}
and substitute \tach\ into \qqq, we find
\eqn\wwa{\eqalign{
\LL_{2n}&=\sum_{l=0}^n a_l^{(n)}\bigg [ \frac{1}{2}
\big( e^{i k_1 \cdot x}-e^{-i k_1 \cdot x}+
(n-1) \big ( e^{ik_2 \cdot x} - e^{-i k_2 \cdot x} \big ) \big )^2  -
\cr
&- (n-1) \sigma ^2 \big ( e^{i k_1 \cdot x}-e^{-i k_1 \cdot x} \big )
\big ( e^{ik_2 \cdot x} -e^{-ik_2 \cdot x} \big ) \bigg ]^l \times
\cr
&\times \bigg [ e^{ik_1 \cdot x}+e^{-ik_1 \cdot x}+(n-1)
\big ( e^{ik_2\cdot x}+e^{-ik_2 \cdot x} \big ) \bigg ]^{2(n-l)}
~ .}}
We expand to linear order in $\sigma ^2$ and drop the $\sigma$-independent
terms to obtain
\eqn\wwb{\eqalign{
\LL_{2n}\sim \sigma^2 \sum_{l=1}^{n} &a_l^{(n)} (n-1) \frac{l}{2^{l-1}}
\bigg [ -e^{i (k_1+k_2) \cdot x}-e^{-i (k_1+k_2) \cdot x} +
e^{i(k_1-k_2) \cdot x} +e^{-i(k_1-k_2) \cdot x} \bigg ]
\times
\cr
&\times \bigg[ e^{i k_1 \cdot x}-e^{-i k_1 \cdot x}+(n-1) \big ( e^{ik_2 \cdot x}
- e^{-i k_2 \cdot x}\big ) \bigg ]^{2l-2}
\times
\cr
&\times \bigg[ e^{ik_1 \cdot x}+e^{-ik_1 \cdot x}+(n-1)\big ( e^{ik_2\cdot x}
+e^{-ik_2 \cdot x} \big ) \bigg ]^{2(n-l)}
~ .}}
Repeated use of the binomial expansion
\eqn\binomial{
(a+b)^m=\sum_{r=0}^m \bigg( {m \atop r} \bigg) a^{m-r} b^r
}
and the identity
\eqn\sigu{
\sum_{i,i'} u_{ii'}= 2(n-1) \sigma^2
}
provides the desired coefficient
\eqn\wwc{\eqalign{
s_{2n} &=\frac{1}{2}\sigma^2 \sum_{l=1}^n \sum_{r_1=0}^{2(n-l)}
 \sum_{r_2=0}^{2(n-l)-r_1}
\sum_{r_3=0}^{r_1} \sum_{s_1=0}^{2l-2}\sum_{s_2=0}^{2l-2-s_1}
 \sum_{s_3=0}^{s_1} a_l^{(n)} \frac{l}{2^{l-1}}
(-)^{n-r_2-r_3-1} (n-1)^{r_1+s_1} \times
\cr
&\times \bigg ( {2(n-l) \atop r_1} \bigg)
\bigg ( {2(n-l)-r_1 \atop r_2} \bigg ) \bigg ( {r_1 \atop r_3} \bigg )
\bigg ( {2l-2 \atop s_1}\bigg )
\bigg ( {2l-2-s_1 \atop s_2}\bigg ) \bigg ( {s_1 \atop s_3} \bigg ) \times
\cr
&\times \bigg [ \delta_{s_2,n-r_2-\frac{1+r_1+s_1}{2}}
\delta_{s_3,\frac{1+r_1+s_1}{2}-r_3}+
\delta_{s_2,n-r_2-\frac{3+r_1+s_1}{2}}
\delta_{s_3,\frac{-1+r_1+s_1}{2}-r_3} +
\cr
&+ \delta_{s_2,n-r_2-\frac{1+r_1+s_1}{2}}
\delta_{s_3,\frac{-1+r_1+s_1}{2}-r_3}+
\delta_{s_2,n-r_2-\frac{3+r_1+s_1}{2}}
\delta_{s_3,\frac{1+r_1+s_1}{2}-r_3} \bigg ]
~ ,}}
with the Kronecker $\delta$ terms in the last parenthesis enforcing the momentum conservation
conditions. For $n=2, 3$ we reproduce the results of sections 3.1 and 3.2
respectively.\foot{In this comparison the appropriate symmetry factors should be reinstated
(1/4 for $n=2$ and $1/20$ for $n=3$).} This provides a trivial check of \wwc. Moreover,
we can show by direct computation of \wwc, or by simple inspection of \wwa\ that the
coefficient of $a^{(n)}_1$, denoted by $d^{(n)}_1$ in \field, vanishes identically for any $n$.
The remaining coefficients $d^{(n)}_l$ are generically non-zero.

For the computation of the tachyon effective action in section 3.4 it was important to establish the
non-vanishing of the overall coefficient $s_{2n}$, or $d^{(n)}$ in \zzz. Substituting the
recursion relation \recurse\ into \wwc\ we find $d^{(n)}$ in terms of the alternating series
\eqn\dn{\eqalign{
&d^{(n)}=\frac{1}{2} \sum_{l=1}^n \sum_{r_1=0}^{2(n-l)} \sum_{r_2=0}^{2(n-l)-r_1}
\sum_{r_3=0}^{r_1} \sum_{s_1=0}^{2l-2}\sum_{s_2=0}^{2l-2-s_1} \sum_{s_3=0}^{s_1}
(-)^{n-1-r_2-r_3} (n-1)^{r_1+s_1} \times
\cr
& \times \frac{(n-1)!(2(n-l))!(2l-2)!}{(n-l)!(l-1)!(2l-1)}
\frac{1}{r_2!(2(n-l)-r_1-r_2)!}\frac{1}{r_3!(r_1-r_3)!} \frac{1}{s_2!(2l-2-s_1-s_2)!}\times
\cr
&\times \frac{1}{s_3! (s_1-s_3)!}
\bigg [ \delta_{s_2,n-r_2-\frac{1+r_1+s_1}{2}} \delta_{s_3,\frac{1+r_1+s_1}{2}-r_3}+
\delta_{s_2,n-r_2-\frac{3+r_1+s_1}{2}} \delta_{s_3,\frac{-1+r_1+s_1}{2}-r_3} +
\cr
&+ \delta_{s_2,n-r_2-\frac{1+r_1+s_1}{2}} \delta_{s_3,\frac{-1+r_1+s_1}{2}-r_3}+
\delta_{s_2,n-r_2-\frac{3+r_1+s_1}{2}} \delta_{s_3,\frac{1+r_1+s_1}{2}-r_3} \bigg ]
~ .}}
We have not been able to find a considerable simplification of this formula that would demonstrate
immediately the non-vanishing of $d^{(n)}$ for all $n$. Instead, we have checked explicitly, using
Mathematica, that up to $n=15$ $\big |d^{(n)}\big |$ is a non-vanishing monotonically
increasing function of $n$. We expect that a similar statement holds for all $n$.

\listrefs
\end